\documentclass[a4paper,twocolumn,floatfix,showpacs,amsmath,amssymb,prb]{revtex4}
\usepackage{graphicx}
\usepackage{dcolumn}
\usepackage{bm}
\usepackage{psfrag}
\usepackage{epsfig}
\usepackage{color}
\newcommand{\upa}{\uparrow}
\newcommand{\doa}{\downarrow}
\begin{document}
\title{Rashba effect and magnetic field in quantum wires}
\author{S.~Debald}
\email{debald@physnet.uni-hamburg.de}
\author{B.~Kramer}
\affiliation{1.~Institut f\"ur Theoretische Physik, Universit\"at
Hamburg,
Jungiusstr.~9, 20355 Hamburg, Germany}
\date{\today}
\begin{abstract}
  We investigate the influence of a perpendicular magnetic field on the
  spectral and spin properties of a ballistic quasi-one-dimensional electron
  system with Rashba effect. The magnetic field strongly alters the spin-orbit
  induced modification to the subband structure when the magnetic length
  becomes comparable to the lateral confinement. A new subband-dependent
  energy splitting at $k=0$ is found which can be much larger than the Zeeman
  splitting. This is due to the breaking of a combined spin orbital-parity
  symmetry.
\end{abstract}
\pacs{85.75.-d, 73.23.Ad, 73.63.Nm, 71.70.Ej}
\maketitle
%
%
\section{\label{sec:intro} Introduction}
The quest for a better understanding of the influence of the electron spin on
the charge transport in nonmagnetic semiconductor nanostructures has
considerably attracted interest during recent years.\cite{ZutFabSar04}
Spin-orbit interaction (SOI) is considered as a possibility to control and
manipulate electron states via gate voltages.\cite{EguBurLos02,RasEfr03} This
has generated considerable research activity, both in theory and experiment,
motivated by fundamental physics as well as applicational aspects. Especially,
SOI induced by the Rashba effect\cite{Ras60,BycRas84} in semiconductor
heterostructures as a consequence of the lack of structure inversion
symmetry\cite{Win03} is important. In these two-dimensional (2D) systems the
Rashba effect leads to spin precession of the propagating electrons. The
possibility to manipulate the strength of the Rashba effect by an external
gate voltage has been demonstrated experimentally.
\cite{NitAkaTakEno97,EngLanSchLue97,Gru00,KogNitAkaTak02} This is the basis of
the spin dependent field-effect-transistor (spinFET) earlier discussed
theoretically by Datta and Das.\cite{DatDas90} Numerous theoretical spintronic
devices have been proposed using interference,
\cite{NitMeiTak99,WanVasPee02,WanVas03,Zue04} resonant tunneling,
\cite{SilRoc99,KogNitTakDat02,GovBoeZueSch02,IonAmi03}
ferromagnet-semiconductor hybrid structures,
\cite{MatEtal0102,LarLunFle02,MirKir02,PalEtal04} multiterminal geometries,
\cite{BulEtal99,KisKim0103,Par04,OheEtal04} and adiabatic
pumping.\cite{GovTadFaz03} Magnetic field effects on the transport properties
in 2D systems with SOI have been investigated theoretically
\cite{TarAve02,WanVas03a,UsaBal04} as well as experimentally
\cite{DasEtal89,MeiEtal04,NitAkaTakEno97,EngLanSchLue97,Gru00,KogNitAkaTak02}.

In order to improve the efficiency of the spinFET the angular distribution of
spin precessing electrons has to be restricted.\cite{DatDas90} Thus, the
interplay of SOI and quantum confinement in quasi-1D
systems\cite{MorBar99,MirKir01,GovZue0204} and quantum Hall edge
channels\cite{PalaGovZueIan04} has been studied. First experimental results on
SOI in quantum wires have been obtained.\cite{SchKnoGuz04} The presence of a
perpendicular magnetic field has been suggested to relax the conditions for
the external confining potential for quantum point contacts. In these systems
a Zeeman-like spin splitting at $k\!=\!0$ has been predicted from the results
of numerical calculations when simultaneously SOI and a magnetic field are
present.\cite{WanSunXin04} The effect of an in-plane magnetic field on the
electron transport in quasi-1D systems has also been
calculated.\cite{CahBan04,PerNesPri04}

In this work, we investigate the effect of a perpendicular magnetic field on
the spectral and spin properties of a ballistic quantum wire with Rashba
spin-orbit interaction. The results are twofold. First, we show that
transforming the one-electron model to a bosonic representation yields a
systematic insight into the effect of the SOI in quantum wires, by using
similarities to atom-light interaction in quantum optics for high magnetic
fields. Second, we demonstrate that spectral and spin properties can be
systematically understood from the symmetry properties. Without magnetic field
the system has a characteristic symmetry property --- the invariance against a
combined spin orbital-parity transformation --- which is related to the
presence of the SOI. This leads to the well-known degeneracy of energies at
$k\!=\!0$. The eigenvalue of this symmetry transformation replaces the spin
quantum number. A non-zero magnetic field breaks this symmetry and lifts the
degeneracy. This magnetic field-induced energy splitting at $k\!=\!0$ can
become much larger than the Zeeman splitting. In addition, we show that
modifications of the one-electron spectrum due to the presence of the SOI are
very sensitive to weak magnetic fields. Furthermore, we find characteristic
hybridisation effects in the spin density. Both results are completely general
as they are related to the breaking of the combined spin-parity symmetry.

This general argument explains the Zeeman-like splitting observed in recent
numerical results.\cite{WanSunXin04}
%
%
\section{\label{sec:model} The Model}
\begin{figure}
\epsfig{file=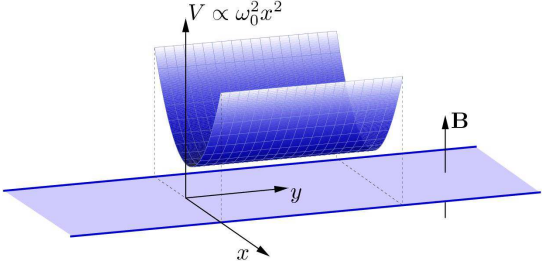,width=0.4\textwidth}
\caption{\label{fig:geometry}(Color online) Model of the quantum wire.}
\end{figure}

%
%
We study a quasi-1D quantum wire with SOI in a perpendicular magnetic field.
The system is assumed to be generated in a 2D electron gas (2DEG) by means of
a gate-voltage induced parabolic lateral confining potential. We assume that
the SOI is dominated by structural inversion asymmetry. This is a reasonable
approximation for InAs based 2DEGs.\cite{KogNitAkaTak02} Therefore, the SOI
 is modelled by the Rashba Hamiltonian,\cite{Ras60,BycRas84}
\begin{equation}\label{eq:model}
H=\frac{(\textbf{p}+\frac{e}{c}\textbf{A})^2}{2m} +
V(x) + \frac{1}{2}g \mu_{\textrm{B}} B \sigma_z -
\frac{\alpha}{\hbar}\big[(\textbf{p}+\frac{e}{c}\textbf{A})\times
\boldsymbol{\sigma}\big]_z,
\end{equation}
where $m$ and $g$ are the effective mass and Land\'e factor of the electron,
and ${\boldsymbol \sigma}$ is the vector of the Pauli matrices.  The magnetic
field is parallel to the $z$-direction (Fig.~\ref{fig:geometry}), and the
vector potential $\textbf{A}$ is in the Landau gauge. Three length scales
characterise the relative strengths in the interplay of confinement, magnetic
field $B$, and SOI,
\begin{equation}
l_0=\sqrt{\frac{\hbar}{m\omega_0}},\quad
l_B=\sqrt{\frac{\hbar}{m\omega_c}}, \quad
l_{\mathrm{SO}}=\frac{\hbar^2}{2m\alpha}.
\end{equation}
The length scale $l_0$ corresponds to the confinement potential
$V(x)=(m/2)\omega_0^2x^2$, $l_B$ is the magnetic length with $\omega_c=eB/mc$
the cyclotron frequency and $l_{\mathrm{SO}}$ is the length scale associated
with the SOI. In a 2DEG the latter is connected to a spin precession phase
$\Delta \theta = L/l_{\mathrm{SO}}$ if the electron propagates a distance $L$.

Because of the translational invariance in the $y$-direction the
eigenfunctions can be decomposed into a plain wave in the longitudinal
direction and a spinor which depends only on the transversal coordinate $x$,
\begin{equation}
\boldsymbol{\Psi}_k(x,y)=e^{iky}
\begin{pmatrix}
\phi_k^{\upa}(x)\\ \phi_k^{\doa}(x)
\end{pmatrix}=:e^{iky}\boldsymbol{\phi}_k(x).
\end{equation}
With this and by defining creation and annihilation operators of a shifted
harmonic oscillator, $a_k^{\dagger}$ and $a_k$, 
which describe the quasi-1D subbands in the case without SOI,
the transversal wavefunction component satisfies
\begin{equation}\label{eq:transverse}
H(k)\,\boldsymbol{\phi}_k(x)=E_k\,\boldsymbol{\phi}_{k}(x),
\end{equation}
for $k$ fixed with the Hamiltonian
\begin{align}\nonumber
\frac{H(k)}{\hbar\omega_0}= &\Omega\left(a_k^{\dagger}a_k+\frac{1}{2}\right)+
\frac{1}{2}\frac{(kl_0)^2}{\Omega^2} \\\label{eq:spin-boson} &+
\frac{1}{2}
\begin{pmatrix}
\xi_1 kl_0 + \xi_2(a_k+a_k^{\dagger})\\\xi_3(a_k-a_k^{\dagger})\\\delta
\end{pmatrix}
\cdot\boldsymbol{\sigma},
\end{align}
the abbreviations
\begin{gather}
  \Omega=\frac{\sqrt{\omega_0^2+\omega_c^2}}{\omega_0}=
\sqrt{1+\left(\frac{l_0}{l_B}\right)^4},
  \\ \xi_1= \frac{l_0}{l_{\mathrm{SO}}}\frac{1}{\Omega},\\
  \xi_2=\frac{1}{\sqrt{2}}\frac{l_0}{l_{\mathrm{SO}}}\left(\frac{l_0}{l_B}
  \right)^ 2 \frac{1}{\sqrt{\Omega}},\\
  \xi_3=\frac{i}{\sqrt{2}}\frac{l_0}{l_{\mathrm{SO}}}\sqrt{\Omega},
\end{gather}
and the dimensionless Zeeman splitting
\begin{equation}
\delta=\frac{1}{2}\left(\frac{l_0}{l_{B}}\right)^2\frac{m}{m_0}g,
\end{equation}
($m_0$ is the bare mass of the electron).  

\begin{figure}
\begin{center}
\epsfig{file=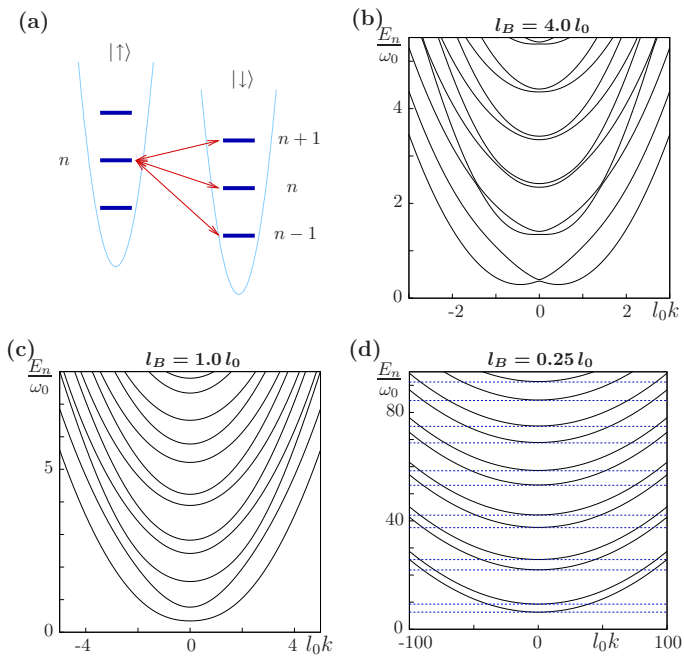,width=0.5\textwidth}
\end{center}
\caption{\label{fig:exact}(Color online) {\bf (a)} Spin-orbit
  induced coupling of subbands with opposite spins in a quantum wire. {\bf
    (b)-(d)} Spectra of a quantum wire with SOI for different strengths of
  perpendicular magnetic field for $l_{\mathrm{SO}}=l_0$ and typical InAs
  parameters: $\alpha=1.0\cdot 10^{-11}\,e$Vm, $g=-8$, $m=0.04\,m_0$. For
  strong magnetic field {\bf (d)} the convergence towards the Jaynes-Cummings
  model (JCM) can be seen (dashed: eigenenergies of JCM).}
\end{figure}

This representation of the Hamiltonian corresponds to expressing the
transverse wavefunction in terms of oscillator eigenstates such that
$a_{k}^{\dagger}a_k$ gives the subband index of the electron which propagate
with longitudinal momentum $\hbar k$. The magnetic field leads to the lateral
shift of the wavefunction and the renormalisation of the oscillator frequency
$\Omega$. Moreover, the effective mass in the kinetic energy of the
longitudinal propagation is changed. The last term in Eq.~(\ref{eq:spin-boson}) describes how the SOI couples the electron's orbital degree of freedom to its spin. Due to the operators $a^{\dagger}_k$ and $a_k$ the subbands corresponding to one spin branch are coupled to the same and nearest neighboring subbands of opposite spin, see Fig.~\ref{fig:exact}a.

Formally, for $k$ fixed Eq.~(\ref{eq:spin-boson}) can be regarded as a simple
spin-boson system where the spin of the electron is coupled to a
mono-energetic boson field which represents the transverse orbital subbands.
This interpretation leads to an analogy to the atom-light interaction in
quantum optics. There, the quantised bosonic radiation field is coupled to a
pseudo-spin that approximates the two atomic levels between which electric
dipol transitions occur. In our model, the roles of atomic pseudo-spin and
light field are played by the spin and the orbital transverse modes of the
electron, respectively.
 
Indeed, in the limit of a strong magnetic field, $l_B \! \ll \! l_0$, and
$kl_0\! \ll\!1$ Eq.~(\ref{eq:spin-boson}) converges against the
exactly integrable Jayne-Cummings model (JCM),\cite{JayCum63}
\begin{equation}
\frac{H_{\textrm{JC}}}{\hbar\omega_c}=a^{\dagger}a+\frac{1}{2}
+ \frac{1}{4}\frac{m}{m_0}g \sigma_ z +
\frac{1}{\sqrt{2}}\frac{l_B}{l_{\mathrm{SO}}}(a\sigma_+
+a^{\dagger}\sigma_-)\,.
\end{equation}
This system is well known in quantum optics. It is one of the most simple
models to couple a boson mode and a two-level system.\cite{AllEbe75} In the
case of the quantum wire with SOI one can show that in the strong magnetic
field limit the rotating-wave approximation,\cite{AllEbe75} wich leads to the
JCM, becomes exact.  This is because for $l_{\textrm{B}}\ll l_0$ and
$kl_0\ll1$ the electrons are strongly localised near the center of the quantum
wire and thus insensitive to the confining potential. In this limit, there is
a crossover to the 2D electron system with SOI in perpendicular magnetic field
for which the formal identity to the JCM has been asserted
previously.\cite{PleZai03}

In this context, it is important to note that the JCM is known to exhibit Rabi
oscillations in optical systems with atomic pseudo-spin and light field
periodically exchanging excitations. Recently, an experimentally feasible
scheme for the production of coherent oscillations in a single few-electron
quantum dot with SOI has been proposed\cite{DebEma04} with the electron's spin
and orbital angular momentum exchanging excitation energy. This highlights the
general usefulness of mapping parabolically confined systems with SOI onto a
bosonic representation as shown in Eq.~(\ref{eq:spin-boson}). Related results
have been found in a 3D model in nuclear physics where the SOI leads to a {\it
  spin-orbit pendulum} effect.\cite{ArvRoz9495}
%
%
\section{\label{sec:symmetry} Symmetry properties}
Without magnetic field it has been pointed out previously that one effect of
SOI in 2D is that no common axis of spin quantisation can be found, see
e.g.~Ref.\cite{GovZue0204}. Since the SOI is proportional to the momentum it
lifts spin degeneracy only for $k\neq 0$. From the degeneracy at $k=0$ a new
binary quantum number can be expected at $B=0$. It can easily be shown that
for any symmetric confinement potential $V(x)=V(-x)$ in Eq.~(\ref{eq:model})
--- which includes the 2D case for $V\equiv 0$ or symmetric multi-terminal
junctions\cite{BulEtal99,KisKim0103,Par04} --- the Hamiltonian is invariant
under the unitary transformation
\begin{equation}\label{eq:spinparity}
U_x=e^{i2\pi \Hat{P}_x \Hat{S}_x/\hbar}=i\Hat{P}_x \sigma_x,\quad
\end{equation}
where $\Hat{P}_{x}$ is the inversion operator for the $x$-component,
$\Hat{P}_x f(x,y)=f(-x,y)$. Thus, the observables $H$, $p_y$ and
$\Hat{P}_x\sigma_x$ commute pairwise. Without SOI, $\Hat{P}_x$ and $\sigma_x$
are conserved separately. With SOI, both operators are combined to form the
new constant of motion $\Hat{P}_x\sigma_x$ which is called {\it spin parity}.
When introducing a magnetic field with a non-zero perpendicular component the
spin parity symmetry is broken and we expect the degeneracy at $k=0$ to be
lifted. As a side remark, by using oscillator eigenstates and the
representation of eigenstates of $\sigma_z$ for the spinor, the Hamiltonian
$H(k)$ in Eq.~(\ref{eq:spin-boson}) becomes real and symmetric. We point out
that in this choice of basis the transformation $U_y=i \Hat{P}_y \sigma_y$ is
a representation of the time-reversal operation which for $B=0$ also commutes
with $H$. However, it does not commute with $p_y$ and no further quantum
number can be derived from $U_y$.\cite{ChaHag98} The effect of the symmetry
$\Hat{P}_x\sigma_x$ on the transmission through symmetric four\cite{BulEtal99}
and three-terminal\cite{KisKim0103} devices has been studied previously.

We recall that the orbital effect of the magnetic field leads to a twofold
symmetry breaking: the breaking of the spin parity $\Hat{P}_x\sigma_x$ lifts
the $k=0$ degeneracy (even without the Zeeman effect) and the breaking of
time-reversal symmetry lifts the Kramers degeneracy. For $B=0$ we can
attribute the quantum numbers $(k,n,s)$ to an eigenstate where $n$ is the
subband index corresponding to the quantisation of motion in $x$-direction and
$s=\pm 1$ is the quantum number of spin parity. For $B\neq0$, due to the
breaking of spin parity, $n$ and $s$ merge into a new quantum number leading
to the nonconstant energy splitting at $k=0$ which will be addressed in the
next Section when treating the spectral properties.

For weak SOI ($l_{\mathrm{SO}}\gg l_0$) one finds in second order that the
spin splitting at $k=0$ for the $n$th subband is
\begin{equation}\label{eq:delta}
\frac{\Delta_n}{\hbar\omega_0}\!
=\delta\!+\!\frac{1}{2}\!\left(\!\frac{l_0}{l_{\mathrm{SO}
} }\!\right)^{\!2}\!\frac{(\Omega- \delta)\chi_1^2\!
-\!(\Omega+\delta)\chi_2^2}{\Omega^2  -\delta^2}\!
\left(\!n+\frac{1}{2}\!\right),
\end{equation}
where $\chi_{1,2}=2^{-1/2}[(l_0/l_B)^2\Omega^{-1/2}\mp \Omega^{1/2}]$.
The first term is the bare Zeeman splitting and the SOI-induced second
contribution has the peculiar property of being proportional to the
subband index. In addition, for weak magnetic field ($l_0\ll l_B$) the
splitting is proportional $B$,
\begin{equation}\label{eq:delta2}
\frac{\Delta_n}{\hbar\omega_0}\! \approx \delta
-\left(\frac{l_0}{l_{\mathrm{SO}}}\right)^2\!
\left(\frac{l_0}{l_B}\right)^2\!
\left(1\!+\! \frac{1}{4}\frac{m}{m_0}g\right)\!\left(n +\frac{1}{2}\right).
\end{equation}
This is expected because by breaking the spin parity symmetry at non-zero $B$
the formerly degenerate levels can be regarded as a coupled two-level system
for which it is known that the splitting into hybridised energies is
proportional to the coupling, i.e.~the magnetic field $B$.

%
%
\section{\label{sec:spectrum} Spectral properties}
Due to the complexity of the coupling between spin and subbands in
Eq.~(\ref{eq:spin-boson}), apart from some trivial limits, no analytic
solution of the Schr\"odinger equation can be expected. We find the
eigenfunctions and energies of the Hamiltonian by exact numerical
diagonalisation. Figures \ref{fig:exact}b\ldots d show the spectra for
different strengths of magnetic field and parameters typical for InAs:
$\alpha=1.0\cdot10^{-11}\,e$Vm, $g=-8$, $m=0.04\,m_0$. We set
$l_{\mathrm{SO}}=l_0$ which corresponds to a wire width $l_0\approx 100\,$nm.

For the case without magnetic field it has been asserted previously that the
interplay of SOI and confinement leads to strong spectral changes like
non-parabolicities and anticrossings when $l_{\mathrm{SO}}$ becomes comparable
to $l_0$.\cite{MorBar99,GovZue0204} In Fig.~\ref{fig:exact}b we find similar
results in the limit of a weak magnetic field. However, as an effect of
non-vanishing magnetic field we observe a splitting of the formerly spin
degenerate energies at $k\!=\!0$. For the Zeeman effect it is expected that
the corresponding spin splitting is constant. In contrast, in
Fig.~\ref{fig:exact}b\ldots d the splitting at $k\!=\!0$ depends on the
subband. This additional splitting has been predicted in
Sec.~\ref{sec:symmetry} in terms of a symmetry breaking effect when SOI and
perpendicular magnetic field are simultaneously present.

\begin{figure}[t]
\begin{center}
\epsfig{file=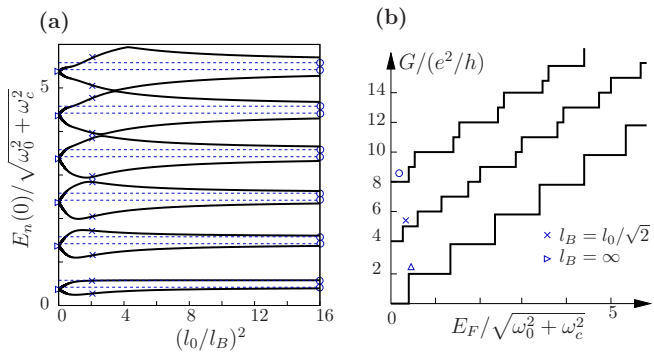,width=0.48\textwidth}
\end{center}
\caption{\label{fig:splitting2}(Color online) {\bf (a)} Magnetic field
  evolution of $E_n(k=0)$ for $l_{\mathrm{SO}}=l_0$ in units of
  $\sqrt{\omega_0^2+\omega_c^2}$. Three different regimes of spin splitting
  can be distinguished (see text). Dashed lines correspond to Zeeman split
  Landau levels. {\bf (b)} Simplified sketch of the ballistic conductance as a
  function of Fermi energy for $(l_0/l_B)^2=0$ (triangles) and $(l_0/l_B)^2=2$
  (crosses) (curves vertically shifted for clarity).}
\end{figure}

Figure \ref{fig:splitting2}a shows the eigenenergies $E_n(k=0)$ for
$l_{\mathrm{SO}}=l_0$ as a function of magnetic field in units of hybridised
energies, $(\omega_0^2+\omega_c^2)^{1/2}$. Three different regimes can be
distinguished. (i)~For small magnetic field ($l_0/l_B\ll1$) the energy
splitting evolves from the spin degenerate case (triangles) due to the
breaking of spin parity. Although the perturbative results
Eq.~(\ref{eq:delta}) cannot be applied to the case $l_{\mathrm{SO}}=l_0$ in
Fig.~\ref{fig:splitting2}a, the energy splitting at small magnetic field and
the overall increasing separation for higher subbands are reminiscent of the
linear dependences on $n$ and $B$ found in Eqs.~(\ref{eq:delta}) and
(\ref{eq:delta2}). (ii)~For $l_0/l_B\!\approx\!1$, the energy splitting is
comparable to the subband separation which indicates the merging of the
quantum numbers of the subband and the spin parity into a new major quantum
number. For higher subbands, the SOI-induced splitting even leads to
anticrossings with neighboring subbands. (iii)~Finally, the convergence to
the JCM implies that the splittings should saturate for large $B$
(Fig.~\ref{fig:exact}d). The dashed lines in Fig.~\ref{fig:splitting2}a show
the energies of the spin-split Landau levels, $E_n/\hbar\omega_0=(l_0/l_B)^2
(n+1/2)\pm \delta/2$ for $l_0=4l_B$, indicating that the SOI-induced energy
splitting is always larger than the bulk Zeeman splitting. At $l_0\approx l_B$
the SOI-induced splitting exceeds the Zeeman effect by a factor 5. This is
remarkable because of the large value of the $g$-factor in InAs.

For our wire parameters the sweep in Fig.~\ref{fig:splitting2}a corresponds to
a magnetic field $B\approx 0 \ldots 1\,$T. Considering the significant
spectral changes due to breaking of spin parity at $l_B\approx l_0$ ($B\approx
70\,$mT) we conclude that the SOI-induced modifications in the wire subband
structure are very sensitive to weak magnetic fields (Fig.~\ref{fig:exact}b,
c). This may have consequences for spinFET designs that rely on spin polarised
injection from ferromagnetic leads because stray fields can be expected to
alter the transmission probabilities of the interface region.

The SOI-induced enhancement of the spin splitting should be accessable via
optical resonance or ballistic transport experiments. In a quasi-1D
constriction the conductance is quantised in units of $ne^2/h$ where $n$ is
the number of transmitting channels.\cite{GlaLesKhmShe88} In the following, we
neglect the influence of the geometrical shape of the constriction and that
for small magnetic fields ($l_0/l_B<0.5$) the minima of the lowest subbands
are not located at $k=0$ (Fig.~\ref{fig:exact}b). In this simplified model we
expect the conductance $G$ to jump up one conductance quantum every time the
Fermi energy passes through the minimum of a subband. Thus, in the case of
spin degenerate subbands, $G$ increases in steps with heights $2e^2/h$
(triangles in Fig.~\ref{fig:splitting2}b).
 
In principle, by sweeping the magnetic field the different regimes discussed
in Fig.~\ref{fig:splitting2}a can be distinguished in the ballistic
conductance. For high magnetic field, the spin degeneracy is broken due to the
Zeeman effect. This leads to a sequence of large steps (Landau level
separation) interrupted by small steps (spin splitting) (circles in
Fig.~\ref{fig:splitting2}b).  As a signature of the SOI we expect increasing
spin splitting for higher Landau levels due to converging towards the JCM
(Fig.~\ref{fig:exact}d). Decreasing the magnetic field enhances the effects of
SOI until at $l_B\approx l_0$ subband and spin splitting are comparable
whereas the Zeeman effect becomes negligible (crosses in
Fig.~\ref{fig:splitting2}b).
%
%
\section{\label{sec:spin}Spin properties}
%
Not only the energy spectra of the quantum wire are strongly affected by the
breaking the spin parity $\Hat{P}_x \sigma_x$. The latter symmetry has also
profound consequences for the spin density,
\begin{equation}\label{eq:spindensity}
\textbf{S}_{n,k}(x):=\psi_{n,k}^{\dagger}\,\boldsymbol{\sigma}\,\psi_{n,
k }. 
\end{equation}
To elucidate this in some detail we start with considering the case
$B=0$.
\subsection{Vanishing magnetic field}
Without magnetic field, the spin parity is a constant of motion. The corresponding symmetry operation Eq.~(\ref{eq:spinparity}) leads to the symmetry property for the wavefunction,
\begin{equation}\label{eq:onsager}
\psi_{n,k,s}^{\uparrow}(x)=s \,\psi_{n,k,s}^{\downarrow}(-x),\quad s=\pm
1,
\end{equation}
where $s$ denotes the quantum number of the spin parity. This symmetry
requires the spin density components perpendicular to the confinement to be
antisymmetric, $S_{n,k,s}^{y,z}(x)=-S_{n,k,s}^{y,z}(-x)$, leading to vanishing
spin expectation values, $\langle\sigma_{y,z}\rangle_{n,k,s}=\int dx
S_{n,k,s}^{y,z}(x)=0$.  We note that using the $\sigma_z$-representation for spinors even leads to zero longitudinal spin {\it density} $S_{n,k,s}^y(x)\equiv 0$ because the real and symmetric Hamiltonian $H(k)$ implies {\em real} transverse wavefunctions independent of the spin parity. Therefore, it is sufficient to consider the $x$- and $z$-components of the spin, only.

For zero magnetic field, it has been pointed out that for large $k$ the spin
is approximately quantised in the confinement direction.\cite{GovZue0204} This
is due to the so-called {\it longitudinal-SOI approximation}\cite{MirKir01}
which becomes valid when the term linear in $k$ in the SOI
[Eq.~(\ref{eq:spin-boson})] exceeds the coupling to the neighboring subbands.
\subsection{Non-vanishing magnetic field}
The perpendicular magnetic field breaks spin parity and thereby leads to a
hybridisation of formerly degenerate states for small $k$. In addition, the
breaking of the symmetry of the wavefunction Eq.~(\ref{eq:onsager}) leads to
modifications of the spin density.

\begin{figure}
\epsfig{file=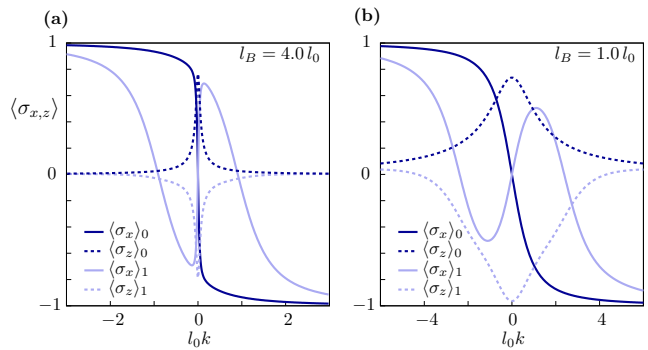,width=0.47\textwidth}
\caption{\label{fig:sigmaexp}Expectation values of spin for the two
lowest eigenstates for $l_{\mathrm{SO}}=l_0$. Solid:
$\langle \sigma_x \rangle_n$, dashed: $\langle \sigma_z\rangle_n$.
\textbf{(a)} weak magnetic field, $l_B=4.0\,l_0$. Hybridisation
of wavefunction at $k\approx 0$ leads to finite $\langle \sigma_z \rangle_n$ component.  \textbf{(b)} Strong magnetic field, $l_B=1.0\,l_0$.}
\end{figure}

In Fig.~\ref{fig:sigmaexp} the expectation value of spin is shown as a function
of the longitudinal momentum for the two lowest subbands.
For weak magnetic field (Fig.~\ref{fig:sigmaexp}a) results similar to the zero
magnetic field case\cite{GovZue0204} are found. For large $k$ the spinor is
effectively described by eigenstates of $\sigma_x$ which concurs with the
longitudinal-SOI approximation. However, for $k\approx 0$ the hybridisation of
the wavefunction leads to a finite value of $\langle \sigma_z \rangle$. This
corresponds to the emergence of the energy splitting at $k=0$ in
Fig.~\ref{fig:exact}b which can be regarded as an additional effective Zeeman
splitting that tilts the spin into the $\sigma_z$-direction --- even without a
real Zeeman effect. This effect becomes even more pronounced for large
magnetic field (Fig.~\ref{fig:sigmaexp}b). Here, for small $k$, the spin of
the lowest subband is approximately quantised in $\sigma_z$ direction. The
spin expectation values in Fig.~\ref{fig:sigmaexp} depend only marginally on the
strength of the Zeeman effect. No qualitative difference is found for $g=0$.
%
%
%
\section{Conclusion}
%
%
%
In summary, the effect of a perpendicular magnetic field on a quasi-1D
electron system with Rashba effect is investigated. It is shown that the
spectral and spin features of the system for small $k$ are governed by a
compound spin orbital-parity symmetry of the wire. Without magnetic field this
{\it spin parity} is a characteristic property of symmetrically confined
systems with Rashba effect and leads to a new binary quantum number which
replaces the quantum number of spin. This symmetry is also responsible for the
well-known degeneracy for $k=0$ in systems with Rashba effect. A non-zero
magnetic field breaks the spin-parity symmetry and lifts the corresponding
degeneracy, thus leading to a novel magnetic field induced energy splitting at
$k=0$ which can become much larger than the Zeeman splitting. Moreover, we
find that the breaking of the symmetry leads to hybridisation effects in the spin
density.

The one-electron spectrum is shown to be very sensitive to weak magnetic
fields. Spin-orbit interaction induced modifications of the subband structure
are strongly changed when the magnetic length becomes comparable to the
lateral confinement of the wire. This might lead to consequences for spinFET
designs which depend on spin injection from ferromagnetic leads because of
magnetic stray fields.

For the example of a quantum wire, we demonstrate
that in the case of a parabolical confinement it is useful to map the
underlying one-electron model onto a bosonic representation which shows for
large magnetic field many similarities to the atom-light interaction in
quantum optics. In Ref.~\cite{DebEma04} this mapping is utilised to predict
spin-orbit driven coherent oscillations in single quantum dots.
\section*{ACKNOWLEDGEMENTS}
This work was supported by the EU via TMR and RTN projects
FMRX-CT98-0180 and HPRN-CT2000-0144, and DFG projects Kr~627/9-1,
Br~1528/4-1. We are grateful to T.~Brandes, T.~Matsuyama and
T.~Ohtsuki for useful discussions.

\end{document}